# Measuring inertial mass with Kibble balance

Rajendra P. Gupta*
*Department of Physics, University of Ottawa, Ottawa, Canada K1N 6N5*

**ABSTRACT**
A Kibble balance measures the *gravitational* mass (weight) of a test mass with extreme precision by balancing the gravitational pull on the test mass against the electromagnetic lift force. The uncertainty in such mass measurement is currently $\sim 1 \times 10^{-8}$. We show how the same Kibble balance can be used to measure the *inertial* mass of a test mass, that too with potentially 50% better measurement uncertainty, i.e., $\sim 5 \times 10^{-9}$. For measuring the inertial mass, the weight of the test mass and the assembly holding it is precisely balanced by a counterweight. The application of the known electromagnetic force accelerates the test mass. Measuring the velocity after a controlled elapsed time provides the acceleration and, consequently, the inertial mass of the accelerated assembly comprising the Kibble balance coil and the mass holding pan. Repeating the measurement with the test mass added to the assembly and taking the difference between the two measurements yields the inertial mass of the test mass. Thus, the extreme precision inertial and gravitational mass measurement of a test mass with a Kibble balance could provide a test of the equivalence principle. We discuss how the two masses are related to the Planck constant and other coupling constants and if the Kibble balance could be used to test the dynamic constants theories in Dirac cosmology.

**Keywords:** Instrumentation; Kibble balance; Dirac cosmology; equivalence principle

## 1. INTRODUCTION

The standard method of measuring the mass of a substance is by comparing it with a standard mass in a balance. Earth's gravity pulls both the masses, and thus, a balance measures the gravitational mass of a substance. Measuring masses are calibrated using a balance and a standardized mass.

By countering a standard weight in a balance on one pan with an electromagnetic force applied from a current-carrying coil under either one of the pans, one can, in principle, determine the current flowing through the coil. The principle was used by Lord Kelvin (Thomson 1888) to realize a working device, the ampere balance, to measure the current flow in a circuit. The precision of the ampere balance was limited by the accuracy of the tedious calculations of the geometric factor used for determining the electromagnetic force generated by the current-carrying nested coils and other conductors (Snow 1939, Driscoll 1958). Kibble (1976) invented a moving mode of the coil to measure the geometric factor of the current-carrying conductors in the balance.

The advent of the quantum Hall effect by Klitzing, Dorda, and Pepper (1980) led to the development of devices with Hall resistance as an integer fraction of $h/e^2$ with $h$ being the Planck constant and $e$ being the elementary charge. When used in conjunction with the voltage precisely measured using the Josephson effect (Josephson 1962), it became possible to measure current with extreme precision. The role of the ampere balance was reversed: Instead of mass measuring current, the resistance and voltage measurements could be used to determine the mass in terms of the Planck constant and vice versa. Since the moving coil mode used electrical power, the balance was generally identified as Watt Balance until it was renamed Kibble Balance to honor Dr. Bryan Kibble (see Schlamminger and Haddad 2019 for a succinct review). Standard Kibble balance measures *gravitational* mass. Alternatively, the gravitational mass can be used to define the Planck constant. Measurement uncertainties achieved are in the range of 10 parts per billion with the Kibble balances at NRC (National Research Council, Canada; Sanchez et al. 2014) and NIST (National Institute of Science and Technology; Possolo et al. 2018).

Cabiali (1991) proposed, and Kibble and Robinson (2014) investigated potential horizontal arrangements of Kibble balance for measuring *inertial* mass and Planck constant and discussed the advantages and problems of such an arrangement. Liu and Wang (2018) proposed an alternative method of measuring the Planck constant using a horizontal configuration for measuring *inertial* mass. They claim that their method has an advantage as some systematic errors can be eliminated in the difference calculation of measurements. They showed an uncertainty improvement by a factor of two over the best Kibble balances primarily because the acceleration due to gravity $g$ is eliminated from the measurement equations. However, their horizontal design for test mass acceleration is very bulky, cumbersome, and hard to realize. Unforeseen problems, such as difficulty in achieving true horizontal movement, friction associated with such movement, and vibration isolation, could introduce errors and uncertainties they did not consider. None of the horizontal devices were ever built to our knowledge.

Here we explore how an existing Kibble balance, such as the NIST-4 (e.g., Haddad et al. 2016, Robinson & Schlamminger 2016, Schlamminger & Haddad 2019), could be used to realize the Planck constant relationship with *inertial* mass rather than with the *gravitational* mass and eliminate the gravitational attraction of Earth from the measurement equation. Fortunately, there is nothing uncertain and unforeseen about the existing Kibble balances in operation. So, their performance under modified operations required for the *inertial* mass mode could be easily extrapolated from the *gravitational* mass mode. Thus, by measuring both the inertial mass and the gravitational mass of a test mass, the Kibble balance alone could be used to test the equivalence principle without requiring an alternative method for determining the inertial mass (Massa, Sasso & Mana 2020; Mana & Schlamminger 2022).

We would also like to study if the Kibble balance can be used to constrain the variability of constants predicted by dynamic constant theories in Dirac cosmology (Dirac 1937), such as that we proposed and tested under various cosmological scenarios (e.g., Gupta 2022, 2023, 2024): $G \sim c^3 \sim h^{3/2}, \sim k^{3/2}$ where $k$ is the Boltzmann constant.

## 2. MEASUREMENTS IN THE GRAVITATIONAL MASS MODE

Following closely the work of Haddad et al. (2016) explaining the Kibble balance principle, one measures the mass of a test mass by balancing the downward gravitational force of the Earth $F_z$ on a mass against the upward electromagnetic force produced by adjusting current $I$ in a coil in a radial magnetic field of flux $\Phi$ with the coil attached to the pan holding the test mass $m$, (Figure 1):

$$F_z = mg = -I\left(\frac{\partial \Phi}{\partial z}\right). \qquad (1)$$

Here $g$ is the local acceleration due to gravity. The derivative of the magnetic flux $\partial \Phi / \partial z$ is determined by measuring the induced voltage $U$ across the coil when the coil is moved in the field with velocity $v_z$:

$$-\frac{\partial \Phi}{\partial z} = \frac{U}{v_z}. \qquad (2)$$

Combining the two equations, we eliminate $\partial \Phi/\partial z$ and get equivalence between the mechanical power and the electrical power:

$$mgv_z = UI. \qquad (3)$$

The precision of mass determination depends on the accuracy of measuring other quantities in Eq. (3).

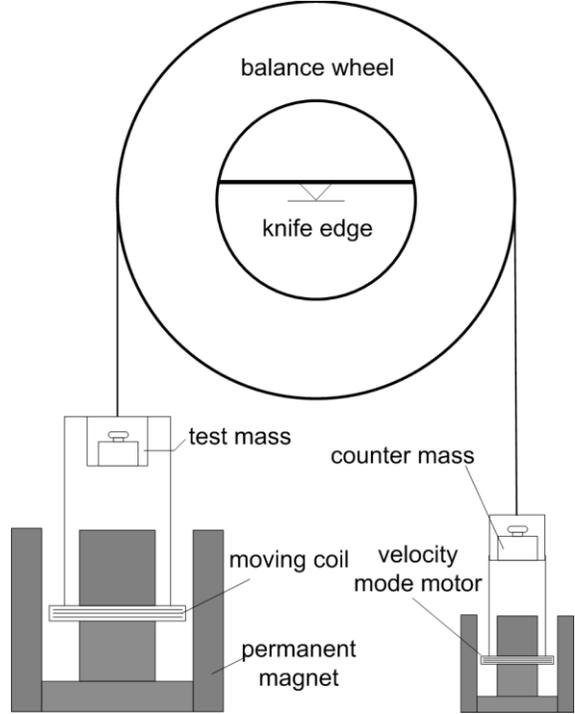

Fig. 1 Schematics of the NIST Kibble balance

The voltage $U$ is measured using the Josephson effect: the small potential difference between two superconductors across a thin insulating barrier driven at a frequency $f$ is $fh/(2e)$ with $e$ the elementary charge and $h$ the Planck constant. A larger voltage is obtained by using $n$ junctions in series: $U = nfh/2e$.

The current $I$ is measured using the quantum Hall effect: when the electric current is confined to flow in a two-dimensional electron system at low temperatures at a high magnetic field, the quotient of the transverse voltage to the current flowing is quantized, i.e., the Hall resistance is an integer fraction of $h/e^2$. Thus, the resistance $R = h/(e^2 p)$, where $p$ is an integer, and

$$I = U/R. \qquad (4)$$

Equation (3) can now be written as

$$mgv_z = \frac{U^2}{R} = \frac{pn^2 f^2 h}{4}. \tag{5}$$

The frequency $f$ is measured as a rational multiple $k_u$ of the standard hyperfine splitting frequency $\Delta v_{Cs}$ of cesium $^{133}$Cs atoms:

$$f = k_u \Delta v_{Cs}. \tag{6}$$

The velocity $v_z$ may be expressed as a fraction $k_v$ of the speed of light $c$:

$$v_z = k_v c. \tag{7}$$

Also, we could write the acceleration due to gravity $g$ as the change in the velocity $v_g$ over time $t_g$ measured in terms of the cesium frequency $\Delta v_{Cs}$. Expressing $v_g$ as a fraction $k_g$ of $c$, and $t_g$ as the inverse of the rational multiple $k_t$ of $\Delta v_{Cs}$, we get:

$$g = \frac{v_g}{t_g} = k_g c \times k_t \Delta v_{Cs}. \tag{8}$$

Combining the above equations, we may write:

$$m = \frac{pn^2}{4}\left(\frac{h\Delta v_{Cs}}{c^2}\right)\left(\frac{k_u^2}{k_g k_t k_v}\right) \equiv m_g. \tag{9}$$

Equation (9) establishes the relationship between the Plank constant and the *gravitational* mass.

## 3. MEASUREMENTS IN THE INERTIAL MASS MODE

In this mode, the total gravitational mass of the coil assembly and the mass-pan (with or without the test mass) on the left side of the balance is precisely balanced with the gravitational mass on the right side pan (within the frictional tolerance of the Balance Wheel knife-edge), Figure 1. It is a non-trivial task but can be achieved, e.g., by in-situ controlled vacuum vapor deposition (if the vacuum of the balance chamber is high enough) or laser erosion techniques. Any electromagnetic force applied would now be on the inertial mass on the left side of the balance. The current in the coil is set such that it results in accelerating all the mass to a set velocity $v_s$ within a set time $t_s$. This current will be different without and with the test-mass $m$ on the mass pan.

According to the momentum definition, force $F$ is the rate of change of momentum $mv$:

$$F_z = m dv_z/dt. \tag{10}$$

Here we have ignored the frictional forces as discussed in Sec. 4. Therefore, using Eq. (1), we may write

$$-\int_0^{t_s} I(t)(d\Phi/dz)dt = \int_0^{v_s} m dv_z.$$

We may now write the two momentum equations without the test mass with subscript $wo$ and with the test mass with subscript $w$ as follows:

$$m_{wo} v_s + \int_0^{t_s}(d\Phi/dz)_{wo} I_{wo}(t)dt = 0, \text{ and} \tag{11}$$

$$m_w v_s + \int_0^{t_s}(d\Phi/dz)_w I_w(t)dt = 0. \tag{12}$$

The derivative of the magnetic flux $\partial\Phi/\partial z$ is again determined by measuring the induced voltage $U$ across the coil when the coil is moved in the field with velocity $v_z$ using Eq. (2):

$$U_{z,wo} = -v_{z,wo}(d\Phi/dz)_{wo}, \text{ and} \tag{13}$$

$$U_{z,w} = -v_{z,w}(d\Phi/dz)_w, \tag{14}$$

The current flow is measured the same way as in the previous section, given by Eq. (4):

$$I_{wo}(t) = U_{wo}(t)/R, \text{ and} \tag{15}$$

$$I_w(t) = U_w(t)/R. \tag{16}$$

Since $m \equiv m_w - m_{wo}$, and $d\Phi/dz$ is assumed time-independent, Eqs. (11) to (16) lead to

$$mv_s = \frac{U_{z,w}}{v_{z,w}}\int_0^{t_s} I_w(t)dt - \frac{U_{z,wo}}{v_{z,wo}}\int_0^{t_s} I_{wo}(t)dt, \text{ or}$$

$$mv_s R = \frac{U_{z,w}}{v_{z,w}}\int_0^{t_s} U_w(t)dt - \frac{U_{z,wo}}{v_{z,wo}}\int_0^{t_s} U_{wo}(t)dt \tag{17}$$

If the currents $I_w(t)$ and $I_{wo}(t)$ are *time-independent*, i.e., applied currents are unaffected by induced currents due to the motion of the coil in the magnetic field, then Eq. (17) simplifies to

$$mv_s = \left(\frac{t_s}{R}\right)\left(\frac{U_{z,w}}{v_{z,w}}U_w - \frac{U_{z,wo}}{v_{z,wo}}U_{wo}\right). \tag{18}$$

Comparing Eq. (18) with Eq. (5), one immediately notices there is no gravitational parameter $g$ involved in the Eq. (18). This equation uses the differences between two quantities. Thus, systematic errors in measuring potentials and velocities will be minimized. Eq. (18) could now be written explicitly to show the $m$ dependence on $h$ and $c$ for comparison with Eq. (9). Following the method in the previous section, we write in terms of constants $k$ subscripted to indicate their association with the respective parameters:

$$t_s = 1/(k_{ts}\Delta v_{Cs}), \tag{19}$$

$$R = h/(e^2 p), \tag{20}$$

$$v_s = k_{vs}c, \qquad (21)$$

$$v_{z,w} = k_{vzw}c, \qquad (22)$$

$$v_{z,wo} = k_{vzwo}c, \qquad (23)$$

$$U_w = nk_w \Delta v_{Cs} h/2e, \qquad (24)$$

$$U_{z,w} = nk_{zw} \Delta v_{Cs} h/2e, \qquad (25)$$

$$U_{wo} = nk_{wo} \Delta v_{Cs} h/2e, \qquad (26)$$

$$U_{z,wo} = nk_{zwo} \Delta v_{Cs} h/2e. \qquad (27)$$

We may now write Eq. (18)

$$m = \frac{pn^2}{4}\left(\frac{h\Delta v_{Cs}}{c^2}\right)\left(\frac{k_{zw}k_w}{(k_{vs}k_{ts})k_{vzw}} - \frac{k_{zwo}k_{wo}}{(k_{vs}k_{ts})k_{vzwo}}\right) \equiv m_i. \qquad (28)$$

Comparing Eq. (9) with Eq. (28), we notice that while the former has four independent $k$-constants, the latter has eight (since $k_{vs}k_{ts}$ in the denominators are the same in the two terms of the last factor). It should be expected as we have to measure the inertial mass of the left side assembly of the balance twice: first without the test mass and second with the test mass. Nevertheless, the operation of the Kibble balance in the inertial mass mode is straightforward. It does not need any modification, except fine-tuning the counterbalancing masses in the right-hand side mass pan.

We may now write

$$\frac{m_g}{m_i} = \left(\frac{k_u^2}{k_g k_t k_v}\right) \Big/ \left(\frac{k_{zw}k_w}{(k_{vs}k_{ts})k_{vzw}} - \frac{k_{zwo}k_{wo}}{(k_{vs}k_{ts})k_{vzwo}}\right). \qquad (29)$$

This ratio must be exactly unity (within the measurement uncertainties) in compliance with the Einstein's weak equivalence principle. Any departure would be in violation of the equivalence principle.

Before we go further, we wish to address two points: (a) In order to achieve the same velocity $v_s$ within the same time $t_s$ for the operations with and without the test mass, one has to know how $U_w$ is related to $U_{wo}$. Assuming the currents $I_w(t)$ and $I_{wo}(t)$ are *time-independent*, one can show that

$$U_w = \left(\frac{m_w}{m_{wo}}\right)\left(\frac{v_{zwo}}{v_{zw}}\right)\left(\frac{U_{zwo}}{U_{zw}}\right) U_{wo}. \qquad (30)$$

Actual value of $U_w$ must be obtained through fine-tuning around this value.
(b) If the currents $I_w(t)$ and $I_{wo}(t)$ are *time-dependent*, then one must use Eq. (17) instead of Eq. (18) for estimating the inertial mass. However, with voltages measured digitally, it should not be very difficult.

As suggested by Schlamminger (2022), the measurements of the inertial mass can be greatly simplified (a) by assuming that $d\phi/dz$ is not measurably different with and without the mass on the pan, and (b) by measuring the velocities for the two cases after the same fixed time for the same applied current. Therefore, $U_w = U_{wo} \equiv U$, $U_{z,w} = U_{z,wo} \equiv U_z$, and $k_{vzw} = k_{vzwo} \equiv kv_z$. Now, if in the velocity mode the velocity is varied until the potential developed across the coil is the same as the applied potential to the coil in the acceleration mode, i.e., $U_z = U$ (as when operating the Kibble balance for measuring the gravitational mass), then we have $k_w = k_{zw} = k_{wo} = k_{zwo} \equiv k_u$. Since the velocities achieved in the acceleration mode are different with and without the test mass on the pan, we take this fact into account by splitting Eq. (21) into two: $v_{sw} = k_{vsw}c$, and $v_{swo} = k_{vswo}c$. Consequently, Eqs. (28) and (29) are modified as follows:

$$m = \frac{pn^2}{4}\left(\frac{h\Delta v_{Cs}}{c^2}\right)\left(\frac{k_u^2}{k_{ts}k_{vz}}\right)\left(\frac{1}{k_{vsw}} - \frac{1}{k_{vswo}}\right) \equiv m_i, \qquad (31)$$

$$\frac{m_g}{m_i} = \frac{1}{k_g}\Big/\left(\frac{1}{k_{vsw}} - \frac{1}{k_{vswo}}\right). \qquad (32)$$

## 4. COMPARISON OF MEASUREMENT UNCERTAINTIES

Physical parameter uncertainty for the two modes of operation, the gravitational mass mode (G-mode) and the inertial mass mode (I-mode) are different. Robinson and Schlamminger (2016) have thoroughly discussed the uncertainties and accuracy of measurements. Values of the uncertainties that are most significant for Kibble balance measurements have been tabulated by Wood et al. (2017) as follows ($\times 10^{-9}$):

| | |
|---|---|
| Mass | 6.44 |
| Alignment | 5.66 |
| Resistance | 5.62 |
| Gravity | 4.87 |
| Various weighing | 3.92 |
| Velocity | 3.06 |
| Type A | 3.04 |
| Voltage measurement | 0.88 |
| **Total** | **12.79** |
| (root mean square sum) | |

*Mass Uncertainty*: Adsorption and desorption of gases and vapors on all surfaces affect their masses. It can be minimized if the masses used on both sides of the balance are identical when operating in the I-mode. Since in I-mode, we take the difference of inertial masses of the assembly with and without the test mass, this uncertainty cancels out and could be ignored.

*Alignment Uncertainty*: It is essentially the same in the I-mode for the Kibble balance operation with and without the test mass and thus can be ignored.
*Resistance Uncertainty*: It is affected by the resistor stability. It is the same for the two modes of operation.
*Gravity Uncertainty*: It is a significant source of uncertainty in the G-mode of the Kibble balance operation as $g$ is affected by natural processes within the Earth, around the Earth, as well as man-made dynamical effects. In the I-mode, it is essentially identical on both sides of the balance and cancels out.
*Various Weighing Uncertainties*: The 2$^{nd}$ order magnetization effect is the largest contributor to this uncertainty. Again, this would mostly cancel out in I-mode since we are taking the difference between the two measurements - with and without the test mass.
*Velocity Uncertainty*: It is measured with the laser interferometry method and remains the same for the two modes of the Kibble balance operation.
*Type A Uncertainty*: It relates to the standard deviation of the mean of data from hundreds of observations. As we expect the total uncertainty determined from several measurements in the I-mode to be about half of its value in the G-mode, we have scaled down the Type A uncertainly accordingly for the I-mode.
*Voltage Measurement*: It should be the same for the two modes.
*Friction*: We have ignored friction in Eq. (10) (Schlamminger 2022). While operating Kibble balance in the G-mode one is only concerned with the static friction, operating it in I-mode we should also consider the kinetic friction. However, since the kinetic friction is significantly smaller than the static friction, and since we are taking difference of the two operations in the I-mode, we expect friction should be of less significance in I-mode than in the G-mode. Nevertheless, in principle it should be possible to explicitly account for the frictional forces by operating Kibble balance for the purpose.
*Center of Mass:* The balance wheel has a center of mass and usually the center of mass is either below or above the knife edge, depending if one wants to have a fast, but unstable or a slow, but stable balance. As we accelerate all the masses we also lift or lower the gravitational center of the wheel. However, since we are doing a difference measurement, this effect is expected to drop out. Nevertheless, one hast to make sure that this is truly the case (Schlamminger 2022).

We can now write down the uncertainties for the I-mode operation of the Kibble balance ($\times 10^{-9}$):

| | |
|---|---|
| Resistance | 5.62 |
| Velocity | 3.06 |
| Type A | 1.52 |
| Voltage measurement | 0.88 |
| **Total** | **6.64** |
| (root mean square sum) | |

Thus, the total estimated uncertainty for the I-mode is about half of the G-mode.

## 5. TESTING VARIABILITY OF COUPLING CONSTANTS

In this section, we will explore if we could test the variability of coupling constants with the Kibble balance. We will do it by examining how the mass $m$ will *apparently* evolve if the coupling constants are evolving with the expansion of the Universe (e.g., Gupta 2022d). Local energy conservation consideration leads to the scaling of different forms of energies since they must be correlated, i.e., gravitational self-energy ~ rest-mass energy ~ thermal energy ~ photon energy ~ electrostatic energy, etc. When distances are measured using the speed of light, one can determine the following relations (Gupta 2022d):

$$-\frac{Gm^2}{r} \sim mc^2 \sim k_B T \sim \frac{hc}{\lambda} \sim \frac{e^2}{4\pi\varepsilon_0 r}. \qquad (33)$$

Here $G$ is the Newton gravitational constant, $m$ is the body mass, $k_B$ is the Boltzmann constant, $T$ is the temperature, $\lambda$ is the photon wavelength, $\varepsilon_0$ is the permittivity of free space, and $r$ is the radius or distance measured with speed light $c$, i.e., $r \sim c$. With the fine structure constant $\alpha$ and the photon frequency $\nu$

$$\alpha = \frac{e^2}{2\varepsilon_0 hc} \text{ and } \frac{hc}{\lambda} = h\nu. \qquad (34)$$

Then,

$$-\frac{Gm^2}{r} \sim mc^2 \sim k_B T \sim h\nu \sim \frac{\alpha hc}{2r}. \qquad (35)$$

Assuming $m, e, \alpha,$ and $\nu$ invariants, this can only be satisfied if

$$G \sim c^3; \; k_B \sim c^2; \; h \sim c^2. \qquad (36)$$

It leads to

$$G = G_0 \mathfrak{f}(a)^3; c = c_0 \mathfrak{f}(a); h = h_0 \mathfrak{f}(a)^2; \text{ etc.} \qquad (37)$$

Here $\mathfrak{f}(a)$ is a function that evolves with the cosmological scale factor $a$ of the expanding Universe. The subscript 0 indicates the current value. Equation (37) does not say anything about the form of $\mathfrak{f}(a)$. It could even be a function that does not evolve with the scale factor $a$. However, the function $\mathfrak{f}(a) = \exp(a^\alpha - 1)$ with $\alpha = 1.8$ has been shown to explain multiple astrophysical observations (e.g., Gupta 2021a,b,c, 2022a,b,c; 2023a,b). Let us see how different quantities in the expressions for the mass $m$ scale with the function $\mathfrak{f}(a)$.

*Gravitational Mass Mode:* The acceleration due to Earth's gravity is $g = MG/R^2$ where $M$ is Earth's mass, and $R$ is its radius (distinct from $R$ used for the resistance above). Since distance is measured using the speed of light, $R$ scales the same as $c$. Therefore, $g \sim \mathfrak{f}^3/\mathfrak{f}^2 \sim \mathfrak{f}^1$. Now any atomic transition-related frequency scales as $cR_\infty$ where the Rydberg constant $R_\infty = m_e e^4 / 8\varepsilon_0^2 h^3 c$ with $m_e$ being the electron mass and $\varepsilon_0$ the permittivity of free space $\sim \mathfrak{f}^{-3}$ (Equation 34). It leads to $R_\infty \sim \mathfrak{f}^{-1}$, and $\Delta\nu_{Cs} \sim \mathfrak{f}^0$. Equating the left and right-hand sides of Equation (8), $k_g$ must scale as $\mathfrak{f}^0$. Equation (9) then yields the *apparent* scaling of $m_g \sim \mathfrak{f}^0$. Thus, the gravitational mass measurement cannot test the variation of the coupling constants.

*Inertial Mass Mode:* The only difference in this mode of the Kibble balance operation compared to the gravitational mass mode operation is that we have no $g$ to consider. Thus all the $k$ parameters in Eq. (28) have no scaling. And since $\Delta\nu_{Cs} \sim \mathfrak{f}^0$ as well as $c^2 \sim h \sim \mathfrak{f}^2$, we derive that $m_i \sim \mathfrak{f}^0$, i.e., the inertial mass measurement is unaffected by the variation of the coupling constants. In other words, the measurement of inertial mass with Kibble balance will not be able to test the coupling constants' variation.

## 6. DISCUSSION AND CONCLUSION

With the high precision of measurement achieved by the Kibble balance and X-ray crystal density (XRCD) method, and the relation of the mass with the Planck constant, CODATA-TGFC (Task Group on Fundamental Constants of the committee on Data for Science and Technology) decided on May 20, 2019, to delink mass calibration from IPK (International Prototype Kilogram) which was the one and the only true kilogram against which all others were measured. The mass measurement is now related to the Planck constant, not to the IPK. Under the new definition, in principle, anyone has the possibility to realize the kilogram independently using an appropriate balance or device.

The new mass definition in terms of the Planck constant is based on the assumption that the Planck constant and other coupling constants do not evolve. Possolo et al. (2018) have discussed in detail the historic stability problem of the Planck constant and its convergence to its present value as measuring techniques improved. The current uncertainty of 10 parts per billion in its value can then be considered similar to the uncertainty of any mass measured with the Planck constant as the reference. As measuring methods and techniques improve, one can expect reduced uncertainty in its measurement.

The measurement technique could affect the measured mass if several constants are concurrently varying. Every measurement technique must therefore be carefully analyzed, as we have done in this paper, to ensure other varying coupling constants do not influence the measured mass. It is worth reiterating this would become even more important if uncertainty in the measurement is improved significantly. Indeed, we have shown that Kibble balance measurements cannot determine the variation of the coupling constants. If we considered the variation of only one coupling constant, say $h$ while keeping all others invariant, we would declare $h$ not varying up to the accuracy of the measurements (see Equations 9 and 28). This would be a false conclusion since variations of several constants cancel out in the mass measurement using Kibble balance. We have to be aware of such false conclusion reached in most studies.

In conclusion, we have shown that the existing Kibble balances could be operated to measure not only the *gravitational* mass but also the *inertial* mass of an object. They can thus test the equivalence principle limited by the uncertainty in the measurements. However, it is incapable of measuring the variation of coupling constants.

## ACKNOWLEDGEMENT

A Macronix Research Corporation research grant partially supported this research. The author is grateful to Dr. Barry Wood of NRC (National Research Council of Canada) for an informed discussion on the limitation of the Kibble balance for measuring the Planck constant at a precision better than ten parts in a billion. He wishes to thank his colleague Dr. Rodrigo Cuzinatto for reviewing the preprint of the paper, and Dr. Stephan Schlamminger of NIST (National Institute of Standards and Technology, U.S.A.) for considering the possibility of testing the variability of constants with the most advanced NIST Kibble balance and reviewing critically the preprint.

## DATA AVAILABILITY

Any data used in this paper is from cited references.

## CONFLICT OF INTEREST

There is no conflict of interest related to the finding reported in this paper.